\documentclass[aps,prd,twocolumn,amsmath,amssymb,nofootinbib,groupedaddress]{revtex4-1}
\usepackage{graphicx}
\usepackage{color}
\usepackage{times}
\usepackage{inputenc}
\usepackage{bm}
\usepackage{ulem}
\usepackage{multirow}
\usepackage{float}
\usepackage{url}
\usepackage{natbib}
\usepackage{subfigure}
\usepackage[dvipsnames]{xcolor}
\usepackage[colorlinks=true,citecolor=blue,urlcolor=blue,linkcolor=blue]{hyperref}
\begin{document}
\preprint{}
\title{Aspects of Rotating Anisotropic Dark Energy Stars}

\author{O. P. Jyothilakshmi}
\email{op\_jyothilakshmi@cb.students.amrita.edu}
\author{Lakshmi J. Naik}
\email{jn\_lakshmi@cb.students.amrita.edu}
\author{V. Sreekanth}
\email{v\_sreekanth@cb.amrita.edu}

\affiliation{Department of Physics, Amrita School of Physical Sciences Coimbatore, Amrita Vishwa Vidyapeetham, India.}

\date{\today}
\begin{abstract}
By employing modified Chaplygin fluid prescription for the dark energy, we construct slowly rotating isotropic and anisotropic dark energy stars. The slow rotation is incorporated via general relativistic Hartle-Thorne formalism; whereas the anisotropy is introduced through Bowers-Liang prescription. We consider both the monopole and quadrupole deformations and present a complete analysis of rotating dark energy stars.
By numerically solving the rotating stellar structure equations in presence of anisotropy, we analyse and quantify various properties of dark energy stars such as mass ($M$), radius, mass deformation, angular momentum ($J$), moment of inertia, and quadrupole moment ($Q$), for three different equation of state parameters. We find that anisotropic slow rotation results in significant deformation of stellar mass and thereby affects other global properties studied. For the values of angular frequencies considered, the effect of anisotropy on the stellar structure is found to be more prominent than that due to rotation. The dimensionless quadrupole moment $QM/J^2$ measuring deviation from a Kerr metric black hole was obtained for anisotropic dark energy stars. We observe that dark energy stars with higher anisotropic strength tend to approach the Kerr solution more closely. We report that our results have considerable agreement with various astrophysical observational measurements.

\end{abstract}

\maketitle


\section{Introduction}
\label{sec:intro}
Compact stars are extremely dense stellar bodies, whose internal composition is not well understood at present. In this regard, there have been many studies assuming various internal matter such as normal nuclear matter, hyperons, quarks, Bose-Einstein condensates, dark matter, dark energy and so on. Among these, compact objects made of elusive dark energy have recently attracted interest~\cite{Chapline:2004jfp,Lobo:2005uf,Chan:2008ui,Ghezzi:2009ct,Yazadjiev:2011sm,Rahaman:2011hd,Beltracchi:2018ait,Sakti:2021mvd,Abellan:2023tft,Panotopoulos:2020kgl,Panotopoulos:2021dtu,Pretel:2023nhf,Pretel:2024tjw,Jyothilakshmi:2024zqn}. Various observations suggest the presence of a hidden component called dark energy, that leads to the accelerated expansion of Universe. Some of the models of dark energy are quintessence, K-essence, tachyon field, cosmological constant, phantom field, dilatonic field, Chaplygin gas etc. Detailed reviews on various such models of dark energy can be seen in Refs.~\cite{Copeland:2006wr,Bamba:2012cp}. In 2004, Chapline~\cite{Chapline:2004jfp} was the first to propose the idea of a dark energy star, which was considered  as an alternative to astrophysically observed black holes. Later, various models of dark energy stars were introduced that include anisotropy effects~\cite{Chan:2008ui}, neutron gas~\cite{Ghezzi:2009ct}, and a mixture of baryonic matter and phantom scalar field~\cite{Yazadjiev:2011sm}. Further, a spherical system that collapses from a state of positive pressure to a state with a dark energy core~\cite{Beltracchi:2018ait} and the stability of a dark energy star with a phantom field~\cite{Sakti:2021mvd} were studied. Analyses of radial pulsations in dark energy stars~\cite{Panotopoulos:2020kgl} and neutron stars with dark energy core~\cite{Pretel:2024tjw} also exist. Recently, the non-radial oscillations in dark energy stars were also considered~\cite{Jyothilakshmi:2024zqn}.
\par
In general, the stellar fluid is assumed to be isotropic, while the fact that compact stars have high densities and strong gravity suggests the possibility of pressure anisotropy within them.
The origin of this anisotropy might be due to the presence of rotation, strong electric and magnetic fields, superfluid cores, phase transition, etc. Anisotropy in compact stars was first proposed by Ruderman in 1972~\cite{Ruderman:1972aj} and two years later, Bowers and Liang~\cite{Bowers:1974tgi} constructed a model for anisotropy and obtained the stellar structure equations. Later, various other models of anisotropy such as Quasi-local \cite{Horvat:2010xf}, Herrera-Barreto \cite{Herrera:2013fja}, and covariant \cite{Raposo:2018rjn} were used to obtain different aspects of compact stars. A review on various relativistic models of anisotropic compact stars can be seen in Ref.~\cite{KUMAR2022101662}. 
\par 
Lately, the effect of anisotropy on various astrophysical scenarios has drawn great interest~\cite{Silva:2014fca,Folomeev:2015aua,Deb:2016lvi,Maurya:2015aea,Pretel:2020xuo,Becerra:2024wku}. The tidal deformability of an anisotropic compact star was studied in Ref.~\cite{Biswas:2019gkw}. Further, the effect of anisotropy on radial~\cite{Horvat:2010xf} and non-radial~\cite{Doneva:2012rd,Curi:2022nnt} oscillations were also analysed. In dark energy stars, the effect of anisotropy was considered within Newtonian approximation by using the generalised Chaplygin model in Ref.~\cite{Abellan:2023tft}.
Recently, the radial stability, moment of inertia and tidal deformability of anisotropic dark energy stars were analysed using proper general relativistic treatment~\cite{Pretel:2023nhf}. Further, in our previous work, it was found that the distinct behaviour of non-radial $f$-mode spectra of anisotropic dark energy stars may help in easy detection by future detectors~\cite{Jyothilakshmi:2024zqn}. In the present work, we intend to study the effect of rotation on properties of anisotropic dark energy stars.   
\par
The effect of rotation on stellar structure of compact stars has been studied using different general relativistic methods such as Hartle-Thorne~\cite{Hartle:1967he,Hartle:1968si}, Butterworth-Ipser~\cite{1976ApJ...204..200B} and Komatsu-Eriguchi-Hachisu~\cite{Komatsu:1989ikr,10.1093/mnras/239.1.153} by various physicists in last few decades~\cite{Kojima:1992ie,Glendenning:1992kd,Yoshida:1997bf,Friedman:1997uh,Gupta:1998dr,Rezzolla:2000dk,Stergioulas:2003yp,Berti:2004ny,Jha2008a,Sotani:2010dr,Laskos-Patkos:2023cts}. Among these, the Hartle-Thorne slow rotation mechanism is widely used in several studies of different types of compact objects. This formalism gives various attributes such as the mass, angular momentum and quadrupole moment of the rotating stars upto second order in angular velocity $\Omega$. Many investigations have used this formalism in the study of gravitational wave sensitive $r$-modes~\cite{Nayyar:2005th,Jha10,Jyothilakshmi:2022hys, Gittins:2022rxs}. In Ref.~\cite{Urbanec:2013fs}, 
this method was used to analyse the quadrupole moments of neutron stars and strange stars. The stellar structure of slowly rotating neutron stars within the scalar tensor~\cite{Sotani:2010dr} and $f(R,T)$~\cite{Murshid:2023xsw} theories of gravity were also looked in. Recently, the properties of slowly rotating isotropic exotic compact objects like Bose-Einstein condensate stars at finite temperature~\cite{Aswathi:2023zzn}, dark matter admixture neutron stars~\cite{Cronin:2023xzc} and dark energy stars~\cite{Panotopoulos:2021dtu} were also studied using the Hartle-Thorne formalism. 
Now, coming to anisotropic compact objects, different aspects of the structure of slowly rotating neutron stars were studied in Ref.~\cite{Pattersons:2021lci}. 
Further, tidal deformability and moment of inertia relation
for an anisotropic neutron star was investigated in Ref.~\cite{Das:2022ell}. The quadrupole moments of neutron stars and strange stars were also studied using the Hartle-Thorne formalism and it was found that both neutron and strange stars show entirely different behaviour~\cite{Urbanec:2013fs}. 
Inclusion of the effect of rotation in anisotropic dark energy star models will be of interest and we aim to take up this study in the present paper.
\par
Using the modified Chaplygin prescription for dark energy, we study the stellar structure of slowly rotating isotropic and 
anisotropic dark energy stars and obtain various parameters such as mass, angular momentum, moment of inertia, quadrupole moment, and tidal deformability. 
The anisotropy is included through the Bowers-Liang model and slow rotation via Hartle-Thorne formalism. 
\par 
The manuscript is organised as follows: In Sec. \ref{sec:EoS}, we discuss the modified Chaplygin EoS for dark energy stars used in our analysis. Next, we present the modified Hartle-Thorne formalism for anisotropic compact stars in Sec. \ref{Sec:rotation}. Finally, we discuss the results in Sec. \ref{sec:results} and present the conclusions drawn in Sec. \ref{sec:summ}. 
\newline
\textit{Notations and conventions}: 
Through out this paper, we take Newton's universal gravitational constant $G$ and velocity of light in vaccum $c$ to be $G=c=1$ and follow the metric convention $g_{\mu\nu}=\textrm{diag}(-1,1,1,1)$. M$_\odot$ denotes mass of the Sun. 

\section{Chaplygin model of Dark Energy}
\label{sec:EoS}

\par
In this section, we present the equation of state used to describe dark energy in our analysis. The Chaplygin gas is one of the well known fluid prescription for dark energy, which could explain the unification of dark matter and dark energy ~\cite{Kamenshchik:2001cp,Bilic:2001cg,Bento:2002ps,Gorini:2002kf,Zhang:2004gc,Xu_2012}. 
This model assumes universe is filled with the particular fluid form of the dark energy. 
For our analysis we use a modified version of the Chaplygin prescription~\cite{Pourhassan:2013sw,Saadat:2013ava}, that takes the following form~\cite{Kahya:2015dpa}
\begin{align}
p=A^2 \rho - \frac{B^2}{\rho}. \label{EOS}
\end{align}
Here, $p$ is the pressure of the fluid, $\rho$ is the energy density and $A$ (no unit) and $B$ (units of energy density) are positive constants. It must be noted that as  the pressure vanishes at the surface of the star, the energy density there takes becomes $\rho_s=B/A$. This along with the causality condition ($v_s^2=dp_r/d\rho<1$) puts constraints on the values of the constant ($A^2<0.5$)~\cite{Pretel:2023nhf}. In the present analysis, following previous studies \cite{Panotopoulos:2021dtu,Panotopoulos:2020kgl,Pretel:2023nhf}, we consider the three sets of values for the constants as given in Table \ref{tab:A-B}. 
\begin{table}[h!]
    \centering
    \begin{tabular}{|c|c|c|c|}
    \hline
         &Set I& Set II& Set III\\
         \hline
         A&$\sqrt{0.4}$&$\sqrt{0.425}$&$\sqrt{0.45}$\\
         \hline
         B ($km^{-2}$)& $0.23\times 10^{-3}$&$0.215\times 10^{-3}$&$0.2\times 10^{-3}$\\
         \hline
    \end{tabular}
    \caption{The values of constants $A$ and $B$ of modified Chaplygin equation of state under consideration.}
    \label{tab:A-B}
\end{table}
\begin{figure}[ht]
    \centering    \includegraphics[width=\linewidth,height=7cm]{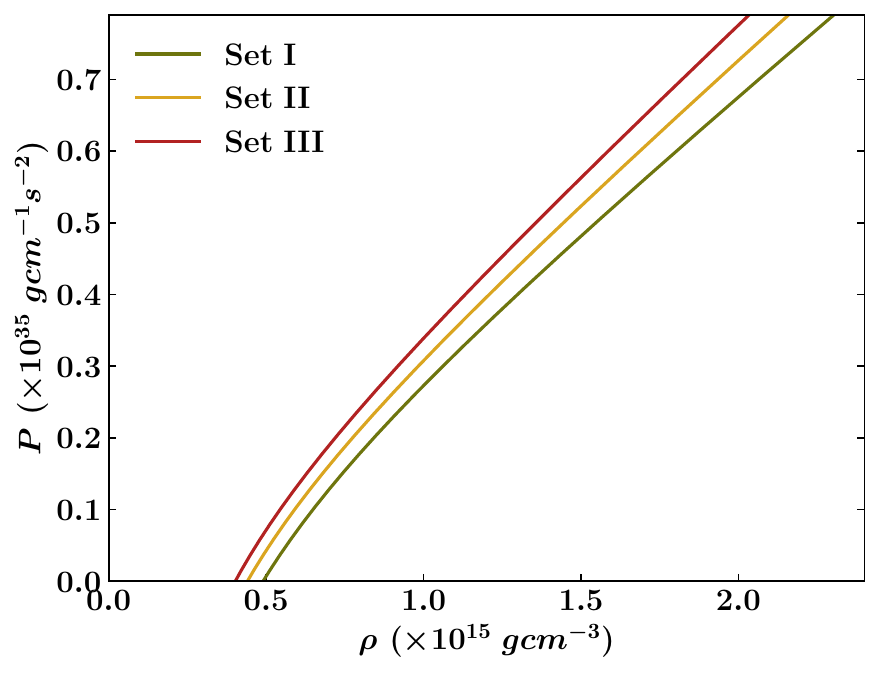}
    \caption{Equation of state of dark energy star that follows the modified Chaplygin prescription, for three parameter sets (See Table \ref{tab:A-B}).}
    \label{fig:eos}
\end{figure}

\par 
In Fig. \ref{fig:eos}, we plot the dark energy equation of state (EoS) for the three sets. We can see that as we increase (decrease) the value of $A$ ($B$), {\it i.e.}, moving from Set I to Set III, the resulting EoS becomes more and more stiffer. 


\section{Slowly rotating anisotropic stars}
 \label{Sec:rotation}

In this section, we present the slowly rotating stellar structure solutions with pressure anisotropy \cite{Pattersons:2021lci} obtained  by modifying the Hartle-Thorne formalism \cite{Hartle:1967he,Hartle:1968si}. Anisotropy in pressure means the pressure in radial and transverse directions are different.  
 The stress-energy tensor $T_{\mu\nu}$ for an anisotropic fluid can be given as
\begin{align}
\label{Tmunu}
        T^{\mu}_{\nu}&=\rho u^\mu u_\nu + q\delta^{\mu}_{\nu} +\sigma k^{\mu}k_{\nu};
\end{align}
where, $\sigma=p-q$ is the anisotropy pressure, $p$ is the radial pressure, $q$ is the tangential pressure, $u^\mu$ is the four velocity of the fluid, and $k_\nu$ denoting the radial vector. The four-vectors obeys the relations $u_\mu u^\mu =-1$,  $k_{\mu}k^{\mu}=1$, and $u^\mu k_{\mu}=0$. For modelling anisotropy we use the Bowers-Liang model given by \cite{Bowers:1974tgi}, 
\begin{align}
    \sigma=-\lambda_{BL}\frac{ r^3}{3}\left(\rho+3p\right)\left(\frac{\rho + p}{r-2m}\right),
\end{align}
with $\lambda_{BL}$ describing the strength of anisotropy. The metric for a spherically symmetric static star in Schwarzschild coordinates, ($t,\, r,\, \theta,\, \phi$) can be written as 
\begin{align}
   ds^{2} &= -e^{2\nu} dt^{2}+e^{ \Lambda} dr^{2}+r^{2}\left(d\theta ^{2}+\sin ^{2}\theta d\phi ^{2}\right).
\end{align}
Here, $\nu(r)$ and $\Lambda(r)$ are the metric functions.
Solving the Einstein field equations for the above metric for an anisotropic fluid results in the modified Tolman-Oppenheimer-Volkoff (TOV) equations describing the stellar structure of the static system, which are given as \cite{Tolman:1939jz,Oppenheimer:1939ne,Bowers:1974tgi} 
\begin{subequations}
  \begin{align}\label{TOV-1}
    \frac{dp}{dr} &=-\frac{(\rho+p)(m+4\pi r^3 p)}{r^2(1-2m/r)}-\frac{2\sigma}{r},\\
    \label{TOV-2}
    \frac{dm}{dr} &= 4\pi r^2 \rho.
  \end{align}   
\end{subequations}
These coupled differential equations reduce to the well known isotropic case (TOV equations), if the anisotropic parameter $\sigma$ is set to zero. For a given EoS, the Eqs. (\ref{TOV-1}) and (\ref{TOV-2}) are integrated from the center to the surface of star where the pressure and mass towards the center are: $p(r=0)=p(\rho_c)$ and $m(r=0)=0$. Also, towards the surface of the star, pressure approaches to zero and mass of the star is obtained as $M=m(r=R)$. Further, we can obtain different masses and radii for all possible values of central densities $\rho_c$.
\par
The pressure $p(r)$ and mass $m(r)$ profiles obtained thus for the static case are used to obtain the solution for a slowly rotating anisotropic star. Here, we make use of the Hartle-Thorne metric \cite{Hartle:1967he,Hartle:1968si}, which is a small perturbation on the metric of a static star:
\begin{align}\label{hmetric}
    ds^2 = &-e^{2\nu}[1+2(h_0+h_2P_2)]dt^2\nonumber\\ 
     &+ r\left[1+2(m_0+m_2P_2)(r-2m)^{-1}\right]/\left(r-2m\right)dr^2 \nonumber\\
    &+ r^2\left[1+2(v_2-h_2)P_2\right]\left[d\theta^2+\sin^2\theta(d\phi-\omega dt)^2\right]\nonumber\\
   & + O(\Omega^3).
\end{align}
Here, $P_2=P_2(\cos\theta)$ is the second order Legendre polynomial and the perturbative terms $h_0,\,m_0,\,h_2,\,m_2,\,p_2,$ and $v_2$ 
are functions of $r$ and are proportional to the square of angular velocity of the star $\Omega^2(r)$. The quantity $\omega$ is the angular velocity of the local inertial frame, given by $\omega=\Omega - \bar{\omega}$. 
For an anisotropic star, the angular velocity relative to the local inertial frame $\bar{\omega}$ is obtained by solving the following second order differential equation \cite{Pattersons:2021lci}
\begin{equation}\label{omega}
    \frac{1}{r^4}\frac{d}{dr}\left(r^4j\frac{d\Bar{\omega}}{dr}\right) +\frac{4}{r}\frac{dj}{dr}\bar{\omega}(1-\frac{\sigma}{\rho+p})=0;
\end{equation}
where, $j =e^{-\nu}{\left(1-2m/r\right)}^{1/2}$. The above differential equation is integrated from the center $(r=0)$ to the surface $(r=R)$ of the star using the boundary conditions: $\bar{\omega} (r=0)=\omega_c$ and $(d\Bar{\omega}/dr)_{r=0} = 0$, where the value of $\omega_c$ is chosen arbitrarily. The angular momentum $J$ and the angular velocity 
$\Omega$ corresponding
to $\omega_c$ are
\begin{equation}
    J=\frac{1}{6}R^4\left(\frac{d\Bar{\omega}}{dr}\right)_{r=R}, \qquad \Omega= \Bar{\omega}(R) + \frac{2J}{R^3};
\end{equation}
and the moment of inertia $I=J/\Omega$ of a slow rotating anisotropic star is given as 
\begin{align}
    I = \frac{8\pi}{3}\int_{0}^{R}\left(\rho + p + \sigma\right)e^{\lambda-\nu}r^4\left(\frac{\bar{\omega}}{\Omega}\right)dr.
\end{align}
 The function $\bar{\omega}(r)$ can be re-scaled to obtain a different value of angular velocity as: ${\Bar{\omega}(r)}_{new} ={\Bar{\omega}(r)}_{old}\left (\Omega_{new}/\Omega_{old}\right)$.
\par
Now, we proceed to obtain the first ($l=0$) and second ($l=2$) order perturbation functions. 
 The first order (spherical) deformation is obtained by solving the mass perturbation factor ($m_0$) and the pressure perturbation factor ($p_0^*$) equations given by \cite{Pattersons:2021lci}
\begin{subequations}
    \begin{eqnarray} \label{HT_m}
        \frac{dm_0}{dr} &=& 4\pi r^2\frac{d\rho}{dp}\left(\rho+p\right)\left(1-\frac{\sigma}{\rho+p}\right)\left(1-\frac{d\sigma}{dp}\right)^{-1}p_0^*  \nonumber\\
        &+& \frac{j^2 r^4}{12} \left(\frac{d\Bar{\omega}}{dr}\right)^2 - \frac{r^3}{3} \frac{dj^2}{dr}\left(1-\frac{\sigma}{\rho+p}\right)\Bar{\omega}^2,\\
        \frac{dp_0^*}{dr} &=& - \label{HT_p}\frac{m_0(1+8\pi r^2p)}{r{(r-2m)}^2} - \frac{4\pi  r^2\left(\rho+p-\sigma\right)}{(r-2m)(1-d\sigma/dp) }p_0^*\nonumber \\ 
        &+& \frac{1}{12}\frac{r^4j^2}{(r-2m)}{\left (\frac{d\Bar{\omega}}{dr}\right)}^2 + \frac{1}{3}\frac{d}{dr}\left(\frac{r^3 j^2 {\Bar{\omega}^2}}{r-2m}\right).
    \end{eqnarray}
\end{subequations}
The above differential equations are solved from the center to the surface of the star and obey the conditions that at the center of the star $m_0=p_0^* =0$. The spherical deformation $\xi_0$ is expressed in terms of the perturbation factor $p_0*$: 
\begin{eqnarray}
    \xi_0(r) &=& -p_0^*\left(\rho+p\right)\left(1-\frac{\sigma}{\rho+p}\right)\left(\frac{dp}{dr}-\frac{d\sigma}{dr}\right)^{-1}.
\end{eqnarray}
Further, the change in gravitational mass $\delta M$ is given by
\begin{equation}
    \delta M = m_0(R) + \frac{J^2}{R^3}.
\end{equation}
Now, in order to obtain the metric functions $v_2$ and $h_2$ describing the second order (quadrupole) deformations, we solve the following differential equations~\cite{Hartle:1968si,Pattersons:2021lci}:
\begin{subequations}
\begin{eqnarray}
    \frac{dv_2}{dr} &=& - 2\frac{d\nu}{dr}h_2 + \left(\frac{1}{r} + \frac{d\nu}{dr}\right)\left[r^3\frac{dj^2}{dr}\bar{\omega}^2\left(\frac{2\sigma}{\rho + p}-\frac{1}{3}\right)\right] \nonumber\\
    &+& \left(\frac{1}{r} + \frac{d\nu}{dr}\right)\left[\frac{1}{6}j^2r^4\left(\frac{d\bar{\omega}}{dr}\right)^2\right],\label{HT-v2}\\
    \frac{dh_2}{dr} &=& - 2\frac{d\nu}{dr}h_2 + \frac{8\pi r (\rho + p)}{r-2m}\left(2\frac{d\nu}{dr}\right)^{-1}h_2 \left(1 - \frac{\sigma}{\rho + p}\right)\nonumber\\
    &\times&\left(1 - \frac{d\sigma}{dp}\right)^{-1} - \frac{r}{r-2m}\left(2\frac{d\nu}{dr}\right)^{-1}\frac{4m}{r^3}h_2\nonumber\\  
    &+& \frac{r^3j^2}{6}\left(\frac{d\bar{\omega}}{dr}\right)^2\Bigg[\frac{d\nu}{dr}r -
   \frac{1}{r-2m}\left(2\frac{d\nu}{dr}\right)^{-1}\Bigg]\nonumber\\
   &-&\frac{4v_2}{r(r-2m)}\left(2\frac{d\nu}{dr}\right)^{-1} - \frac{r^3\bar{\omega}^2}{3}\frac{d\nu}{dr}\frac{dj^2}{dr}\bar{\omega}^2\left(1 - \frac{6\sigma}{\rho + p}\right)\nonumber\\
   &-& \frac{r^2\bar{\omega^2}}{6(r-2m)}\left(\frac{dj^2/dr}{d\nu/dr}\right)\left(1 - \frac{\sigma}{\rho + p}\right)\left(1- \frac{d\sigma}{dp}\right)^{-1}.\label{HT-h2}
\end{eqnarray}
\end{subequations}
The solutions of the coupled differential equations given above must be regular near the center of the star ($r=0$) and can be expressed as
\begin{eqnarray}
h_2^{P} &\to& \mathcal{A} r^2,\qquad
v_2^{P} \to \mathcal{B} r^{4};\label{BC-P}
\end{eqnarray}
where constants $ \mathcal{A}$ and $\mathcal{B}$ are related by
\begin{equation}
\mathcal{B} + \frac{2 \pi}{3} (\rho_c + 3 P_c) \mathcal{A} =  -\frac{4 \pi}{3} (\rho_c + P_c) (j_c\bar{\omega}_c)^2.
\end{equation}
Here, subscript '$c$' is used to denote the center of the star. The particular solutions $h_2^P$ and $v_2^P$ are obtained by integrating Eqs.~(\ref{HT-v2}) and (\ref{HT-h2}) from the center to the surface of the star using the conditions given by Eq.~(\ref{BC-P}). Then, we solve the homogeneous differential equations~\cite{Hartle:1968si}
\begin{subequations}
\begin{eqnarray}
\frac{dv_2^{H}}{dr}&=&-2\frac{d\nu}{dr}h_2^{H}, \label{eq:v2H} \\
    \frac{dh_2^H}{dr} &=& - 2\frac{d\nu}{dr}h_2^H - \frac{r}{r-2m}\left(2\frac{d\nu}{dr}\right)^{-1}\frac{4m}{r^3}h_2^H\nonumber\\
    &+& \frac{4\pi r (\rho + p)h_2^H}{(r-2m)(d\nu/dr)}\left(1 - \frac{\sigma}{\rho + p}\right)\left(1 - \frac{d\sigma}{dp}\right)^{-1}\nonumber\\  
      &-&\frac{4v_2^H}{r(r-2m)}\left(2\frac{d\nu}{dr}\right)^{-1}.\label{eq:h2H} 
\end{eqnarray}
\end{subequations}
Near the center ($r=0$) of the star these equations obey the following conditions:
\begin{eqnarray}
h_2^{H} &\to& \mathcal{B}^H r^2, \qquad v_2^{H} \to  - \frac{2 \pi}{3} (\rho_c + 3 P_c)  \mathcal{B}^H r^4.
\end{eqnarray}
Here, we take $\mathcal{B}^H$ to be unity and then integrate Eqs. (\ref{eq:v2H}) and (\ref{eq:h2H}) from the center to the surface of the star. The internal solutions of Eqs. (\ref{HT-v2}) and (\ref{HT-h2}) are expressed as a linear combination of both particular ($v_2^P$ and $h_2^P$) and homogeneous solutions ($v_2^H$ and $h_2^H$) as
\begin{eqnarray}
h_2 &=& h_2^{P} + \mathcal{K}_1 h_2^{H},\\
v_2 &=& v_2^{P} + \mathcal{K}_1 v_2^{H};
\end{eqnarray}
where $\mathcal{K}_1$ is a constant. Outside the star, $h_2$ and $v_2$ take the form
 \begin{eqnarray}
 h_2 &=& J^2 \left( \frac{1}{m r^3} + \frac{1}{r^{4}}  \right) + \mathcal{K}_2 \mathcal
 Q_2^2\left(\frac{r}{m} -1  \right), \\
 v_2 &=& - \frac{J^2}{r^4} + \mathcal{K}_2 \frac{2m}{[r(r-2m)]^{1/2}}
 \mathcal Q_2^1\left(\frac{r}{m} -1 \right);
 \end{eqnarray}
 where $\mathcal{K}_2$ is a constant and the $\mathcal Q_a^b$ are associated Legendre functions of the second kind. The constants $\mathcal{K}_1$ and $\mathcal{K}_2$ can then be obtained by matching the internal values for $v_2$ and $h_2$ to the external values at the surface. Finally, we obtain the mass quadrupole moment $Q$ of a rotating star given by
\begin{equation}
    Q = \frac{8}{5}\mathcal{K}_2M^3 + \frac{J^2}{M}.\label{QM}
\end{equation}
The quadrupole moment $Q$ can be written with respect to the Kerr solution as a dimensionless quantity, $\bar{q}=QM/J^2$ and can be used to describe the deviation of the Hartle-Thorne metric from the Kerr black hole metric, for which the Kerr factor tends to unity. 
\par 
Further, we proceed to calculate the mass and radius of the star under rotation by obtaining 
the quadrupole deformation $\xi_2$, which is expressed as 
\begin{eqnarray}
    \xi_2(r) &=& -p_2\left(\rho+p-\sigma\right)\left(\frac{dp}{dr}-\frac{d\sigma}{dr}\right)^{-1};
\end{eqnarray}
where, $p_2 = -(h_2+ \frac{1}{3}e^{-2\nu}r^2\bar{\omega}^2)$ is the quadrupole pressure perturbation factor. Now, the mass ($M_{rot}$) and radius ($R_{rot}$) of a slowly rotating star in the Hartle-Thorne formalism are given by  
\begin{subequations}
\begin{eqnarray}
    M_{rot} &=&  M + m_0(R) + \frac{J^2}{R^3},  \label{24a}\\
   R_{rot} &=& R+\xi(r,\theta) =R + \xi_0(R) + \xi_2(R)P_2 \label{24b}.
\end{eqnarray}
\end{subequations}
Here, $M$ and $R$ are the mass and radius of a static star obtained from TOV equations for a given central density. We note that, towards the pole and equator of the star, the values of $\theta$ are $0$ and $\pi$ respectively. Therefore, the polar and equatorial radii are defined respectively as 
\begin{eqnarray}
    R_p &=& R + \xi_0(R) + \xi_2(R),\\
    R_e &=& R + \xi_0(R) - \xi_2(R)/2.
\end{eqnarray}
The expressions for various stellar profiles, such as energy density ($\rho_{r}$) and radial ($p_{r}$), tangential ($q_r$) and anisotropic ($\sigma_r$) pressures of slowly rotating anisotropic stars are given respectively as 
\begin{subequations}
 \begin{eqnarray}   
    \rho_{r} &=& \rho - \xi(r,\theta)( d\rho/dr),\\
    p_{r} &=& p - \xi(r,\theta)( dp/dr),\\
    q_{r} &=& q - \xi(r,\theta)( dq/dr),\\
    \sigma_{r} &=&  \sigma - \xi(r,\theta)( d\sigma/dr).
\end{eqnarray}   
\end{subequations}
\par
We make use of this formalism to obtain the stellar configurations of slowly rotating dark energy stars that obey the modified Chaplygin EoS.  
\begin{figure*}[ht]
\centering
\subfigure[]{\includegraphics[width=0.48\linewidth,height=7cm]{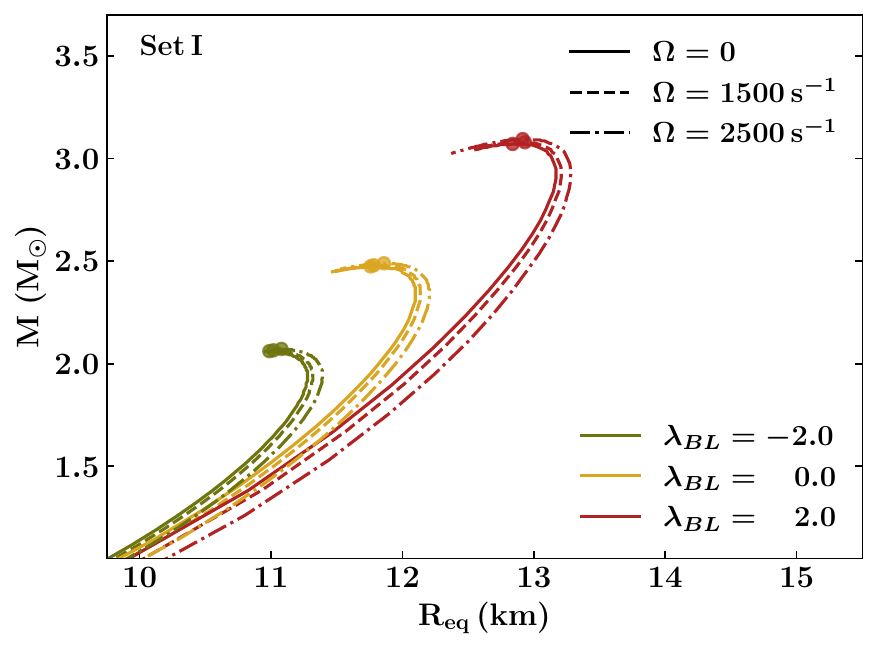}\label{fig:MR-1}}\quad
\subfigure
[]{\includegraphics[width=0.48\linewidth,height=7cm]{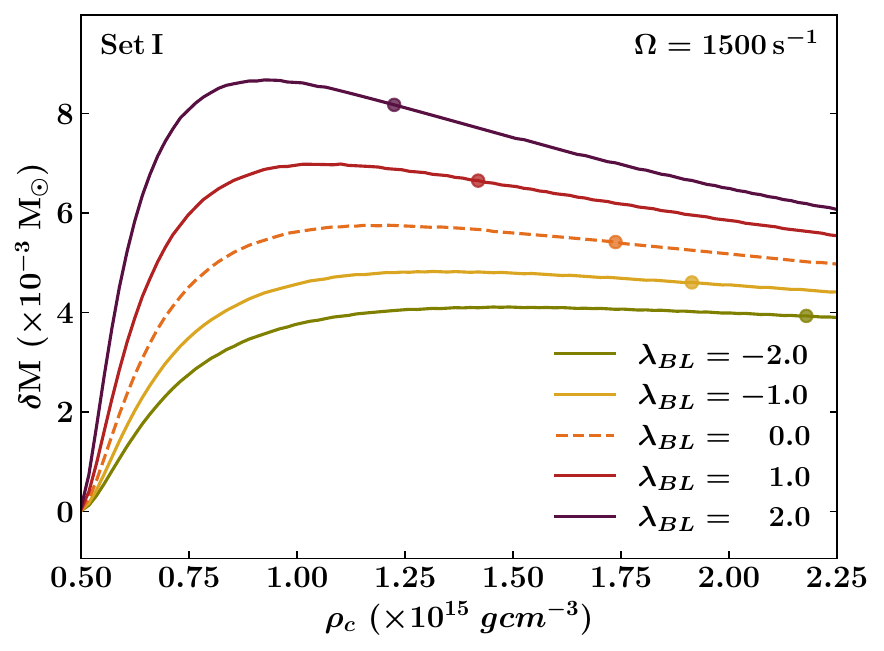} \label{fig:delM-Ec}}
\caption{(a) Stellar mass $M$ as a function of equatorial radius $R_{eq}$ with angular velocity $\Omega=0,\,1500$ and $2500$ s$^{-1}$ and (b) the mass correction $\delta M$ as a function of central density $\rho_c$ with $\Omega=1500$ s$^{-1}$ for dark energy stars by varying the anisotropic parameter $\lambda_{BL}$ from $-2$ to $2$ for EoS Set I (see Table \ref{tab:A-B}).}
\end{figure*}
\begin{figure*}[ht]
\centering
\subfigure[]{\includegraphics[width=0.48\linewidth,height=7cm]{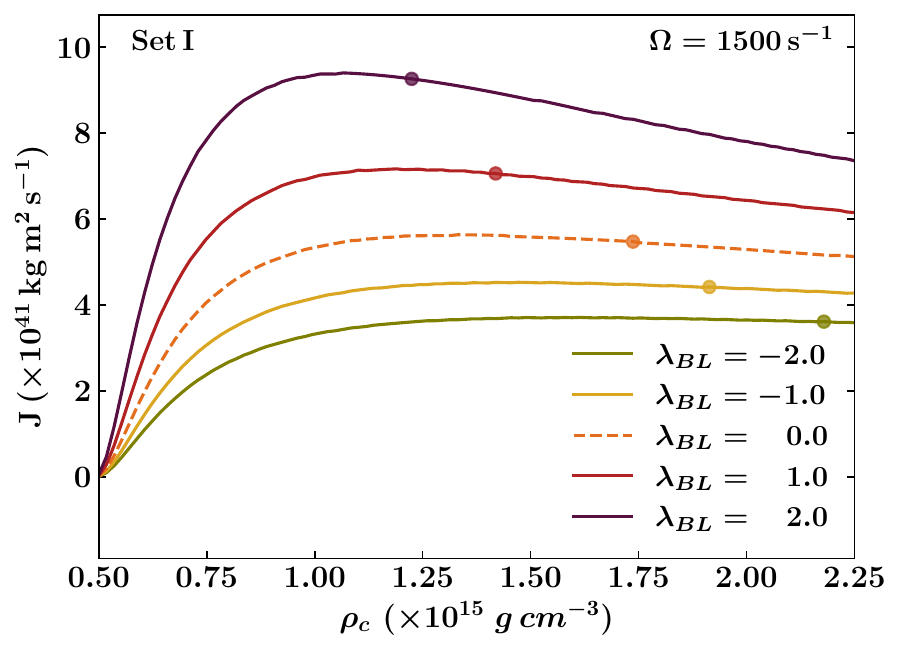} \label{fig:J-Ec}}\quad
\subfigure
[]{\includegraphics[width=0.48\linewidth,height=7cm]{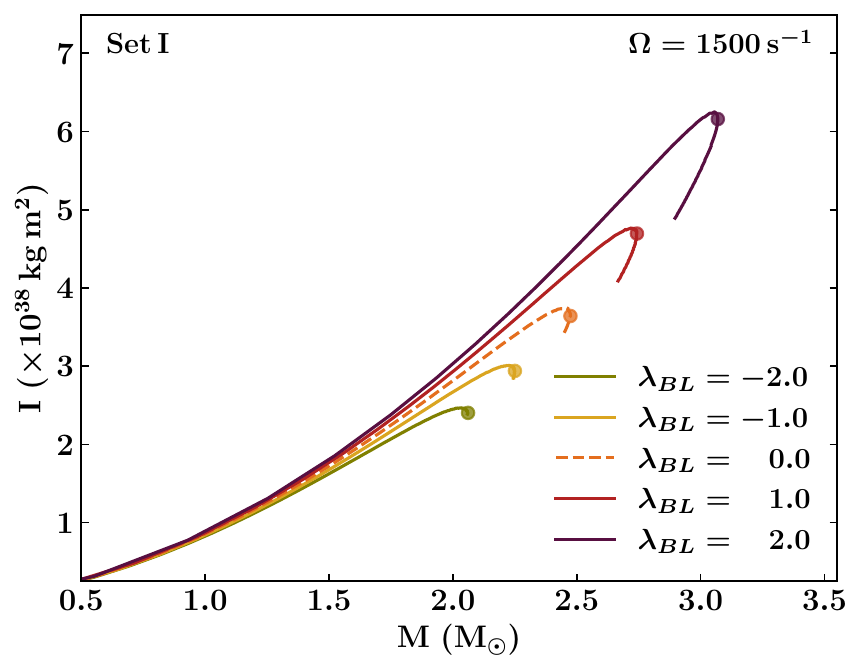}\label{fig:I-M}}
\caption{(a)  Angular momentum $J$ as a function of central density $\rho_c$ and (b) Moment of inertia $I$ as a function of stellar mass $M$ for an angular velocity $\Omega=1500\,s^{-1}$ with $\lambda_{BL}$ values ranging from $-2$ to $2$ for EoS Set I.}
\label{fig:I-J}
\end{figure*}
\begin{figure*}[ht]
\subfigure[]{\includegraphics[width=0.48\linewidth,height=7cm]{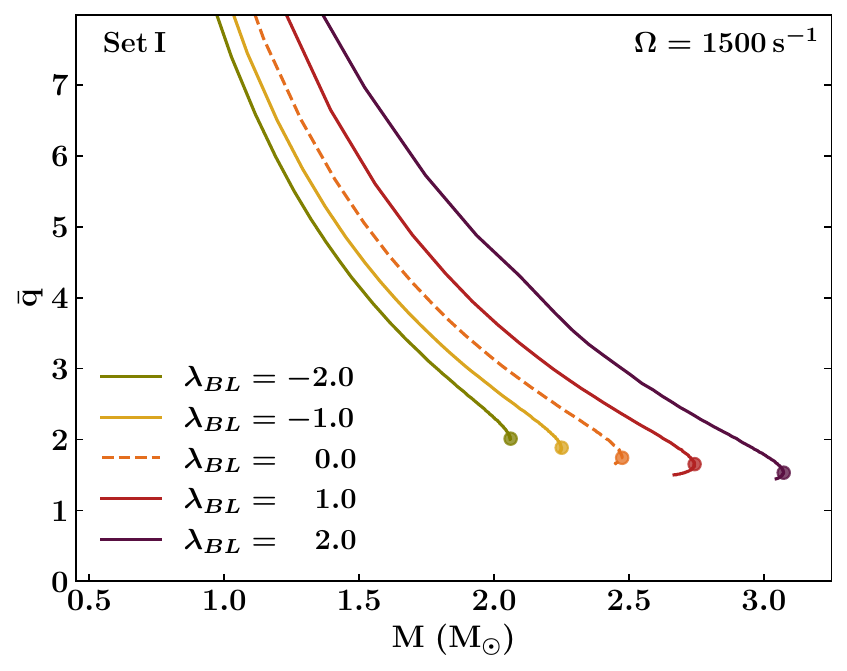}\label{fig:QD}}\quad
\subfigure
[]{\includegraphics[width=0.48\linewidth,height=7cm]{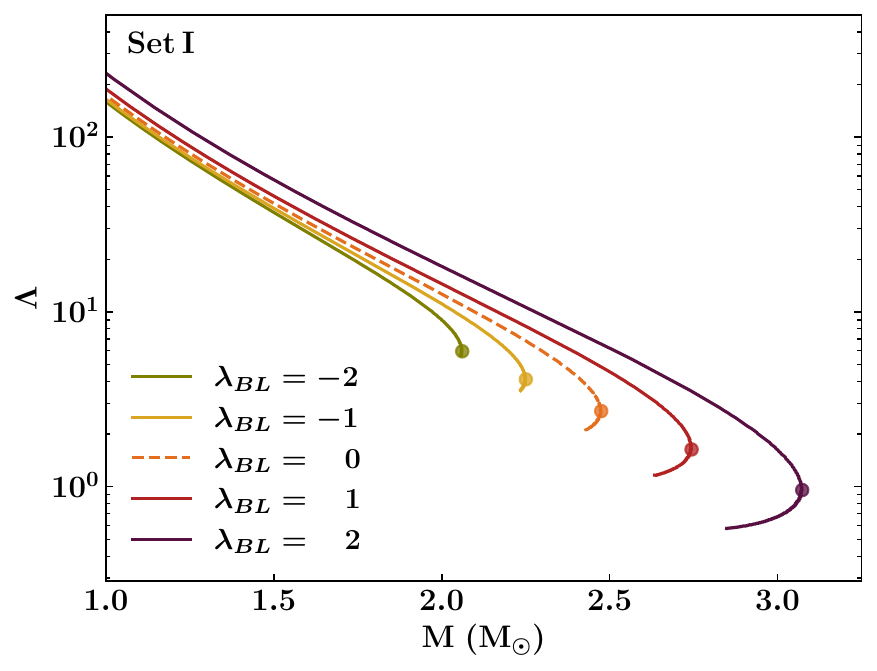} \label{fig:LM}}
\caption{(a) The Kerr factor $\bar{q}=QM/J^2$ and (b) tidal deformability $\Lambda$ as a function of stellar mass $M$ of anisotropic dark energy stars obeying EoS Set I.}
\end{figure*}
\begin{figure}[ht]
    \centering
    \includegraphics[width=\linewidth,height=7cm]{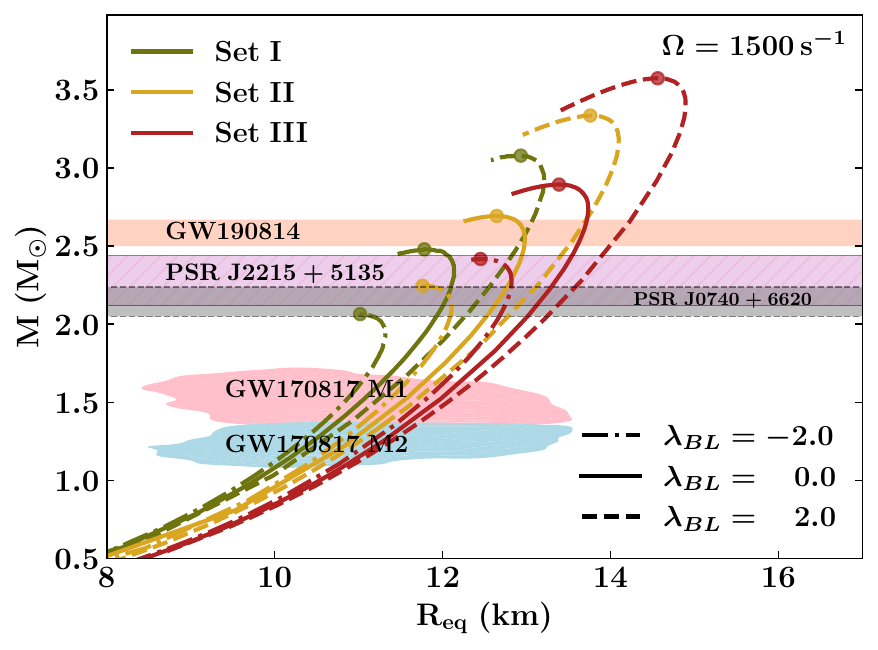}
    \caption{Mass-radius curves of slowly rotating anisotropic dark energy stars for angular velocity $\Omega=1500\,s^{-1}$ for the three different parameter sets discussed in Sec. \ref{sec:EoS} (see Table \ref{tab:A-B}) with $\lambda_{BL}= -2,\,0,$ and $2$. The horizontal bands are observational measurements from GW190814 events~\cite{Abbott2020} and pulsars, PSR J2215+5135 ($M = 2.27_{-0.15}^{+0.17} M_\odot$) \cite{Linares:2018ppq} and PSR J0740+6620 ($2.14_{-0.09}^{+0.10} M_\odot$ ($68.3\% $ credible)) \cite{NANOGrav:2019jur}. The observational constraints from GW 170817 event~\cite{LIGOScientific:2018cki} are also shown.}
    \label{fig:MR-2}
\end{figure}
\begin{figure*}[ht] 
\centering
\subfigure[]{\includegraphics[width=0.48\linewidth,height=7cm]{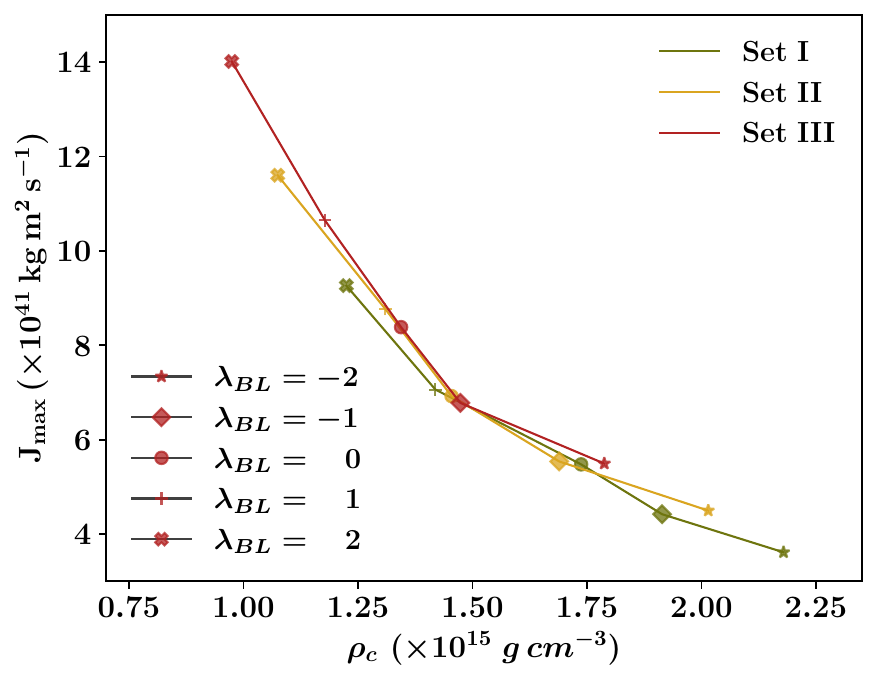}\label{fig:J-Ec-2}}\quad
\subfigure
[]{\includegraphics[width=0.48\linewidth,height=7cm]{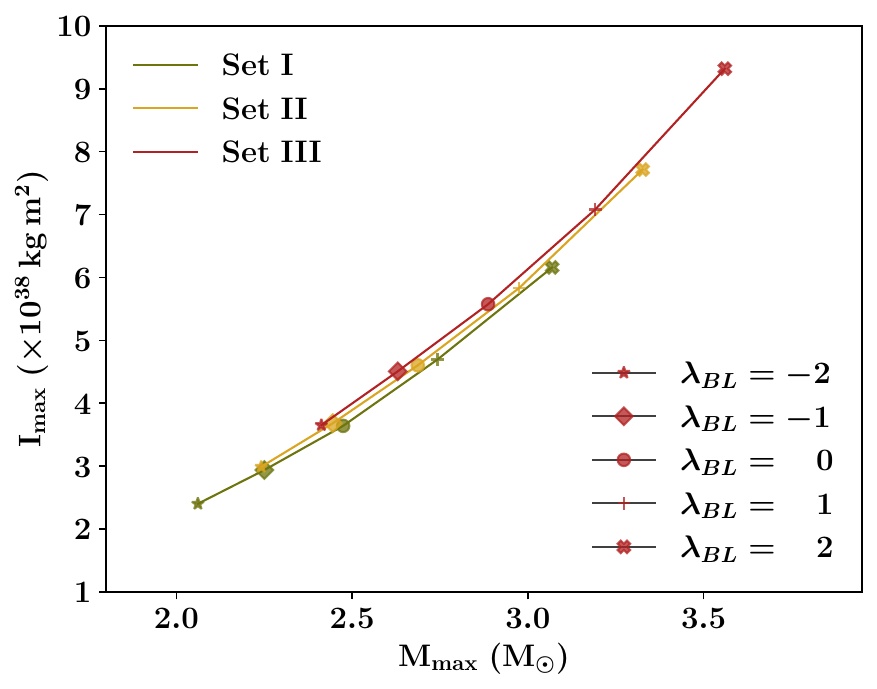} \label{fig:I-M-2}}
\caption{(a) Angular momentum $J$ as a function of central density $\rho_c$ and (b) Moment of inertia $I$ as a function of stellar mass $M$ for an angular velocity $\Omega=1500\,s^{-1}$  for the three different parameter sets discussed in Sec. \ref{sec:EoS} by varying $\lambda_{BL}$ from $-2$ to $2$.}
\label{fig:I-J-2}
\end{figure*}
\begin{figure*}[ht] 
\centering
\subfigure[]{\includegraphics[width=0.48\linewidth,height=7cm]{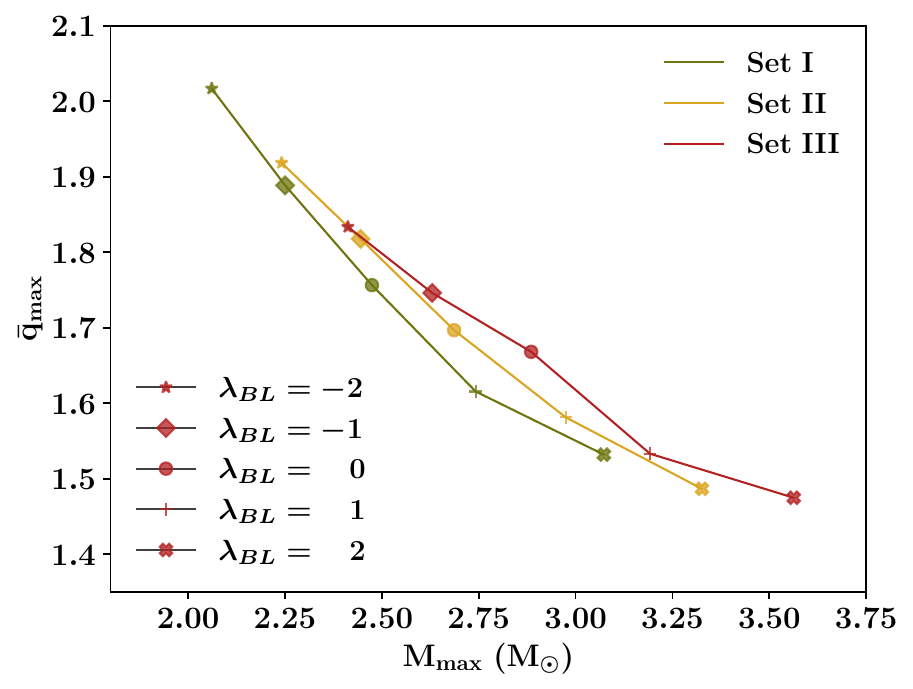}\label{fig:QM-2}}\quad
\subfigure
[]{\includegraphics[width=0.48\linewidth,height=7cm]{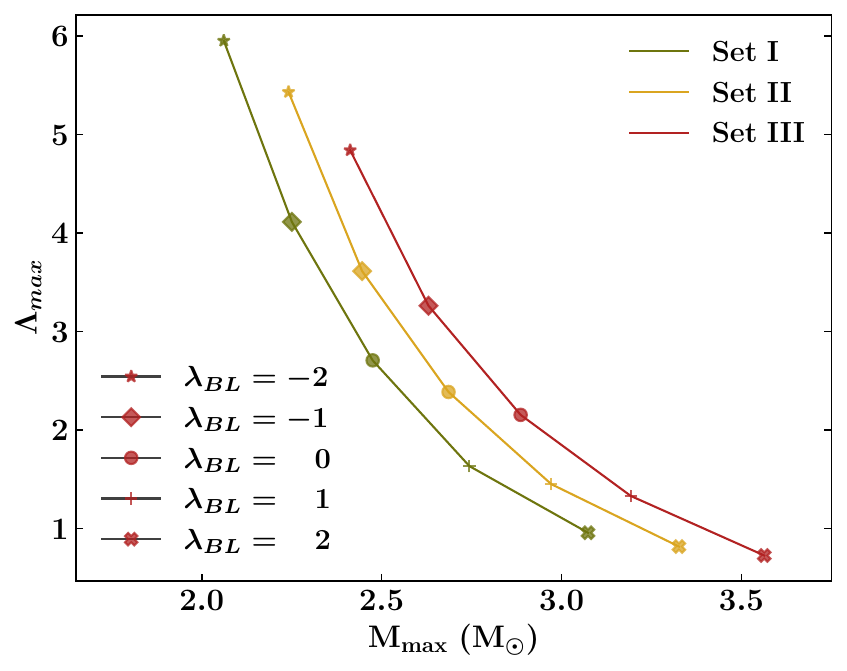} \label{fig:LM-2}}
\caption{(a) Kerr factor ($\bar{q}_{max}$) and (b) tidal deformability ($\Lambda_{max}$) as a function of maximum mass $M_{max}$ for the three different parameter sets discussed in Sec. \ref{sec:EoS} with $\lambda_{BL}= -2,\,-1,\,0,\,1,\,2$.}
\label{fig:I-J-2}
\end{figure*}
\section{Results and discussions}
\label{sec:results}

Here, we present the stellar properties of slowly rotating anisotropic dark energy stars that obey the modified Chaplygin EoS given by Eq.~(\ref{EOS}), for three different parameter sets given in Table~\ref{tab:A-B}. In our analysis, the effect of anisotropy is incorporated using the Bowers-Liang model. We follow the modified Hartle-Thorne formalism discussed in Sec.~\ref{Sec:rotation} to obtain the mass and radius of rotating anisotropic dark energy stars. First, we numerically obtain the solutions for static case by solving the coupled differential equations, Eqs. (\ref{TOV-1}) and (\ref{TOV-2}). These solutions are then used to obtain the stellar properties of slowly rotating dark energy stars by numerically integrating the differential equations corresponding to first (Eqs. (\ref{HT_m}) and (\ref{HT_p}))  and second (Eqs. (\ref{HT-v2}) and (\ref{HT-h2})) order perturbations. In all the plots, we mark the maximum mass points using dots. 

\par 
 In Fig. \ref{fig:MR-1}, we plot the mass of anisotropic dark energy stars as a function of stellar radius for different angular velocities $\Omega=(1500,\,2500)$ s$^{-1}$. We vary the strength of anisotropy, $\lambda_{BL}$ from $-2$ to $2$ for equation of state (EoS) Set I parameters discussed in Sec. \ref{sec:EoS}. The stellar structure of a static dark energy star shows an overall increment (decrement) if we increase (decrease) the value of $\lambda_{BL}$ from zero. This is inline with the previous studies~\cite{Pretel:2023nhf,Jyothilakshmi:2024zqn}. With rotation, the stellar structure shows an overall increment in masses and radii; while only a minimal increment is found in the maximum masses and corresponding radii. The equatorial radii corresponding to a $1.5\,M_\odot$ dark energy star rotating with angular velocties $\Omega=1500\,(2500)$ s$^{-1}$ are $10.85\,(10.94)$, $11.06\,(11.17)$, and $11.25\,(11.44)$ km for $\lambda_{BL}=-2,\,0,$ and $2$ respectively. The corresponding values of central density are $7.6,\,6.4$ and $5.7$ $\times 10^{14}$ g cm$^{-3}$ respectively. The maximum masses (corresponding radii) of rotating dark energy stars show only a minimal deviation from the static case ($\Omega=0$) and their values are approximately $2.07$ M$_\odot$ ($11.08$ km), $2.48$ M$_\odot$ ($11.86$ km) and $3.09$ M$_\odot$ ($12.91$ km) respectively for $\lambda_{BL}=-2,\,0$ and $2$. We find that the maximum masses and corresponding radii increase (decrease) for a positive (negative) deviation of $\lambda_{BL}$ from zero (isotropic case), as observed for the static case~\cite{Jyothilakshmi:2024zqn}. 
\par
Next, we plot the mass correction $\delta M$ due to rotation as a function of central density $\rho_c$ for an anisotropic dark energy star by varying $\lambda_{BL}$ from $-2$ to $2$ in Fig.~\ref{fig:delM-Ec}.
For a fixed value of $\lambda_{BL}$, the $\delta M$ value shows a gradual increase with an increase in $\rho_c$ until it attains a certain value and when $\rho_c$ is increased further, $\delta M$ remains almost constant for profiles with $\lambda_{BL}\leq 0$ and decreases gradually for profiles with $\lambda_{BL}> 0$. A significant increase of mass is found for a dark energy star rotating with an angular velocity ($\Omega=1500$ s$^{-1}$). For a central density $\rho_c=8\times 10^{14}$ gcm$^{-3}$, the values of mass deformation $\delta M $ with $\lambda_{BL}=-2,\,-1,\,0,\,1,\,$ and $2$ are ($3.07,\,3.84,\,4.90,\,6.37,$ and $8.45$) $\times 10^{-3}$ M$_\odot$ respectively. We find that the effect of rotation causes high (low) deformation for a positive (negative) increment of $\lambda_{BL}$. The values of mass correction $\delta M$ corresponding to the maximum mass configuration of dark energy stars are ($3.93,\,4.60,\,5.41,\,6.65,\,8.11$) $\times 10^{-3}$ M$_\odot$ for $\lambda_{BL}=(-2,\,-1,\,0,\,1,\,2)$ respectively. The corresponding central density $\rho_c$ values are ($2.18,\,1.91,\,1.73,\,1.42,\,1.25$) $\times 10^{15}$ g cm$^{-3}$ respectively. We note that the stellar configurations with higher anisotropy values attain the maximum mass for a smaller value of $\rho_c$, which reduces the possible range of $\rho_c$ values that gives stable dark energy profiles. We find that the inclusion of anisotropy results in a substantial change in the mass of a slowly rotating anisotropic dark energy star.
\par
In Fig. \ref{fig:J-Ec}, we plot the angular momentum $J$ as a function of central density $\rho_c$ for rotating anisotropic dark energy stars with an angular velocity $\Omega=1500$ s$^{-1}$. The values of angular momentum corresponding to the maximum mass ($J_{max}$) for $\lambda_{BL}=-2,\,-1,\,0,\,1,\,$ and $2$ are ($3.61,\,4.42,\,5.47,\,7.06$, and $9.26$) $\times 10^{41}$ kg m$^{2}$ s$^{-1}$ respectively. We find that the values of angular momentum increase (decrease) when the $\lambda_{BL}$ value is increased (decreased) from zero. In Fig. \ref{fig:I-M}, we plot the moment of inertia $I$ as a function of stellar mass for a rotating anisotropic dark energy star with an angular velocity $\Omega=1500$ s$^{-1}$. We find that the deviation of $I$ due to anisotropy is more prominent in the high mass region, while it shows only a minimal change in the low mass region. The values of moment of inertia corresponding to the maximum mass ($I_{max}$) for dark energy stars are ($2.40,\,2.94,\,3.64,\,4.69,$ and $6.16$) $\times 10^{38}$ kg m$^{2}$ with $\lambda_{BL}=-2,\,-1,\,0,\,1,\,$ and $2$ respectively. 
\par
 We numerically obtain the quadrupole moment $Q$ using Eq. (\ref{QM}).
In Fig. \ref{fig:QD}, we plot the dimensionless Kerr factor 
$\bar{q}=QM/J^2$ as a function of the stellar mass $M$ for Set I EoS by varying $\lambda_{BL}$ from $-2$ to $2$. We find that the Kerr factor $\bar{q}$ has a high value in the low mass region and as the mass increases, the $\bar{q}$ value decreases and approaches unity. The values of quadrupole moment corresponding to the maximum stellar mass ($\bar{q}_{max}$) are $2.01,\,1.88,\,1.74,\,1.65,$ and $1.53$ respectively for $\lambda_{BL}=-2,\,-1,\,0,\,1,$ and $2$. We find that the $\bar{q}_{max}$ corresponding to the positive (negative) values of $\lambda_{BL}$ tends more (less) towards the Kerr solution. The $\bar{q}_{max}$ values lie between $2-1.5$ for $\lambda_{BL}$ values ranging from $-2$ to $2$, which is almost similar to the range obtained for isotropic neutron and strange stars in Ref.~\cite{Urbanec:2013fs}. 
\par
Next, we plot the tidal deformability $\Lambda$ as a function of stellar mass $M$ for anisotropic dark energy stars by varying $\lambda_{BL}$ from $-2$ to $2$ in Fig. \ref{fig:LM}. 
The dimensionless tidal deformability is given by $\Lambda=2k_2C^{-5}/3$, where $C=M/R$ is the compactness of the star and $k_2$ is the tidal Love number. We obtain the Love number $k_2$ by following the prescription given in Refs.~\cite{Hinderer2008,Hind2010}.
The $\Lambda$ values show only a minimal variation from the isotropic case in the low mass region and the deviation is prominent in the higher mass region. We also note that the $\Lambda$ values are high (low) for positive (negative) values of $\lambda_{BL}$. The values of $\Lambda$ corresponding to $1.40$ M$_\odot$ are $49,\,53,\, 59,\, 60,$ and $76$ for $\lambda_{BL}=-2,\,-1,\,0,\,1,$ and $2$ respectively. These values are in agreement with the limit provided by GW 170817 ($\Lambda_{1.4}\leq 800$)~\cite{LIGOScientific:2017vwq}. 
\par 
We now proceed to study the impact of EoS parameters on rotation and anisotropy in the system by varying the positive constants $A$ and $B$ (see Table. \ref{tab:A-B}) of the modified Chaplygin EoS. We plot the mass-radius relation of a dark energy star rotating with an angular velocity $\Omega=1500\;s^{-1}$ for three different sets of $A$ and $B$ values by varying the anisotropy parameter $\lambda_{BL}$ from $-2$ to $+2$ in Fig. \ref{fig:MR-2}. The maximum mass of dark energy stars with $\lambda_{BL}=(-2,0,2)$ are ($2.065,\,2.480,\,3.077$), ($2.246,\,2.694,\,3.337$) and ($2.418,\,2.895,\,3.572$) M$_\odot$ for Sets I, II, and III respectively. The equatorial radii corresponding to them are ($11.02,\,11.78,\,12.89$), ($11.76,\,12.64,\,13.76$) and ($12.44,\,13.34,\,14.53$) km respectively. We note that for different values of $\lambda_{BL}$, Set III gives the highest maximum mass, since a stiffer EoS results in higher maximum mass.  

\par 
In Fig.~\ref{fig:MR-2}, we also compare our results with some of the recent observational limits obtained from gravitational wave events (GW190814~\cite{Abbott2020}, GW170817~\cite{LIGOScientific:2018cki}) and milli-second pulsars (PSR J2215+5135 ($M = 2.27_{-0.15}^{+0.17} M_\odot$) \cite{Linares:2018ppq} and PSR J0740+6620 ($2.14_{-0.09}^{+0.10} M_\odot$ )  \cite{NANOGrav:2019jur}). We find that our results for rotating anisotropic dark energy stars for all the three sets of EoS parameters lie well within the bound given by the GW170817. The maximum masses of $\lambda_{BL}=-2$ are also consistent with the PSR J2215+5135 and PSR J0740+6620. Further, the masses of dark energy stars with $\lambda_{BL}=2$ obey the limit imposed by the GW190814 for all three parameter sets considered, while for stars with $\lambda_{BL}=0$, only set II and III obey this limit.

\par

In Fig. \ref{fig:J-Ec-2}, we plot the angular momentum corresponding to the maximum mass ($J_{max}$) and analogous central density $\rho_c$ of a slowly rotating dark energy star with angular velocity $\Omega=1500\,s^{-1}$ by varying the $\lambda_{BL}$ for three different parameter sets.  
We find that the $J_{max}$ value increases when we increase the $\lambda_{BL}$ value from $-2$ to $2$. The central density corresponding to the maximum mass is highest for $\lambda_{BL}=-2$ and lowest for $\lambda_{BL}=2$. Therefore, for positive increament of $\lambda_{BL}$ value the star attains the maximum mass configuration with comparitively smaller value of central density. The values of $J_{max}$ for Sets I, II, and III are ($3.61$, $4.42$, $5.48$, $7.06$, $9.26$), ($4.50$, $5.54$, $6.92$, $8.77$, $11.60$) and ($5.49$, $6.78$, $8.38$, $10.65$, $14.01$) $\times 10^{41}$ kg m$^2$ s$^{-1}$ respectively with $\lambda_{BL}=(-2,\,-1,\,0,\,1,\,2)$. The central densities corresponding to the maximum masses are ($2.18$, $1.91$, $1.73$, $1.42$, $1.22$), ($2.01$, $1.69$, $1.45$, $1.31$, $1.07$) and ($1.79$, $1.47$, $1.34$, $1.18$, $0.97$) $\times 10^{15}$ g cm$^{-3}$ respectively. 
\par 
We plot the moment of inertia $I_{max}$ as a function of maximum mass by varying $\lambda_{BL}$ of dark energy stars rotating with an angular velocity $\Omega=1500\;s^{-1}$ for different EoS parameter sets in Fig. \ref{fig:I-M-2}. The values of $I_{max}$ for Sets I, II, and III are ($2.40,\,2.94,\,3.64,\,4.70,\,6.16$), ($2.99,\,3.68,\,4.60,\,5.83,\,7.71$) and ($3.65,\,4.51,\,5.58,\,7.08,\,9.32$) $\times 10^{38}$ kg m$^{2}$ respectively with $\lambda_{BL}=(-2,\,-1,\,0,\,1,\,2)$. We find that the values of moment of inertia for anisotropic dark energy stars with $1.5\,M_\odot$ lie in the range of $(1.7-2.1)\times 10^{38}$ kg m$^2$. We also note that, EoS Set III gives the highest value for $J_{max}$ and $I_{max}$. We note that the order of magnitude of values of $J$ and $I$ for dark energy stars are in agreement with the results given in Refs.~\cite{Panotopoulos:2021dtu,Pretel:2023nhf}. 
\par
Further, in Fig. \ref{fig:QM-2}, we plot the Kerr factor corresponding to the maximum mass ($\bar{q}_{max}$) for different parameter sets by varying $\lambda_{BL}$. The values of $\bar{q}_{max}$ for Sets I, II, and III are ($2.01$, $1.88$, $1.74$, $1.65$, $1.53$), ($1.9$, $1.81$, $1.70$, $1.58$, $1.50$) and ($1.84$, $1.75$, $1.64$, $1.53$, $1.47$) respectively with $\lambda_{BL}=(-2,\,-1,\,0,\,1,\,2)$. The value of $\bar{q}_{max}$ is lowest for $\lambda_{BL}=2$ for all three sets of parameters discussed here. We also plot the tidal deformability corresponding to the maximum mass ($\Lambda_{max}$) by varying $\lambda_{BL}$ for different parameter sets in Fig. \ref{fig:LM-2}. The values of $\Lambda_{max}$ for Sets I, II, and III are ($5.95$, $4.11$, $2.71$, $1.63$, $0.96$), ($5.43$, $3.61$, $2.38$, $1.45$, $0.82$) and ($4.84$, $3.26$, $2.15$, $1.33$, $0.73$) respectively with $\lambda_{BL}=(-2,\,-1,\,0,\,1,\,2)$. 

\section{Summary and Conclusions}
\label{sec:summ}
We have studied the  global properties of slowly rotating isotropic and anisotropic dark energy stars obeying modified Chaplygin EoS. We have used the general relativistic Hartle-Thorne slow rotation approximation method for incorporating the effects of rotation. By including both the monopole and quadrupole deformations, we carried out a complete analysis of slowly rotating dark energy stars.  
We investigated and quantified the mass-radius relation, mass correction, angular momentum, moment of inertia, quadrupole moment and tidal deformability of rotating anisotropic dark energy stars. For our analysis, we made use of three different EoS parameter sets and varied anisotropy parameter $\lambda_{BL}$ from $-2$ to $+2$. 
\par
Firstly, we studied the stellar configurations for different angular velocities by varying the strength of anisotropy $\lambda_{BL}$ for a single set of EoS parameters, $A$ and $B$. We found that for a fixed value of $\lambda_{BL}$, the effect of rotation on maximum mass and corresponding radius is minimal. However, the overall structure of a rotating dark energy star shows significant deviation from the static case. We also obtained the mass correction as a function of central density by varying the values of $\lambda_{BL}$. It was found that the mass correction increases (decreases) for a positive (negative) deviation from the isotropic case ($\lambda_{BL} = 0$). We report that the mass correction due to anisotropy is high compared to the correction obtained by rotation. Also, the maximum mass occurs at smaller central densities for more positive values of $\lambda_{BL}$.  
Further, we have analysed the angular momentum and moment of inertia for the system. The angular momentum of the star shows an overall increment (decrement) for positive (negative) values of $\lambda_{BL}$. It was observed that the effect of anisotropy on moment of inertia is more visible in high mass regions. We also studied the quadrupole deformation and obtained
the quadrupole moments with respect to Kerr solution. It was found that the dark energy stars with higher
anisotropy value approach more towards the Kerr solution. Further, the tidal deformability of the dark energy star was studied by varying $\lambda_{BL}$ and our analysis shows that effect of anisotropy is prominent in the high mass region. The obtained tidal deformability values are consistent with the constraints set by the GW170817 event.
\par
Next, we have studied the stellar properties by varying the EoS parameters. With a stiffer EoS, we obtained stellar configurations with higher maximum masses and corresponding equatorial radii, for any fixed $\lambda_{BL}$ value. We have also compared our results with various astrophysical observations of milli-second pulsars (PSR J2215+5135 and PSR J0740+6620) and GW events (GW190814, GW170817) and considerable agreement was found. The effect of EoS parameters was studied on the angular momentum, moment of inertia, quadrupole moment and tidal deformability. 
It was found that the values of Kerr factor for maximum mass profiles decrease with the stiffness of EoS and lie in the range of $1.47-2.01$, which could be well approximated by the Kerr metric.

\section*{Acknowledgements}
L. J. N. acknowledges the Department of Science and Technology, Govt. of India for the INSPIRE Fellowship.

\bibliography{DE-aniso-rot}	
\end{document}